# Improvement of Sliding Mode Control Strategy Founded on Cascaded Doubly Fed Induction Generator Powered by a Matrix Converter

Amar Maafa
Department of Electrical Engineering
Akli Mohand Oulhadj University
Bouira, Algeria
ommafa@univ-bouira.dz

Hacene Mellah
Department of Electrical Engineering
Akli Mohand Oulhadj University
Bouira, Algeria
h.mellah@univ-bouira.dz

Kaci Ghedamsi
Laboratoire de Maîtrise des Energies Renouvelables
Bejaia University, Algeria
kghedamsi@yahoo.fr

Djamel Aouzellag
Laboratoire de Maîtrise des Energies Renouvelables
Bejaia University, Algeria
aouzellag@hotmail.com

**Abstract-The current paper presents a Sliding Mode Controller (SMC) for indirect field-oriented Cascaded Doubly Fed Induction Generator (CDFIG) powered through a Matrix Converter (MC). The proposed SMC employs a continuous control strategy to accomplish free chattering fractional-order sliding-mode control and to ensure that the control of the first DFIG stator's reactive and active power is separated. An MC is used to control the current provided to the second stator of the CDFIG as an alternative to standard voltage source inverters. The two MCs are controlled via Space-Vector Pulse-Width Modulation (SVPWM) and Indirect Field Oriented Control (IFOC). The proposed Wind Power Generation System (WPGS) is used with the purpose to ensure Maximum Power Point Tracking (MPPT) sensing under various disturbance variables such as turbulent wind. The simulation results prove the efficiency and robustness of the proposed method.**

## I. INTRODUCTION

The current interest in wind energy comes from the need to develop less expensive, clean, and sustainable WPGSs [1], while reducing maintenance cost and increasing availability and production [2], hence, the need for the research regarding electromechanical converters of high reliability [2]. Many WTs are equipped by DFIGs [3] to produce electrical energy [4-9]. Furthermore, brushes and slip-rings, are fitted with wound-rotor induction machines, which may pose maintenance issues. In addition, because of its similar characteristics, a CDFIG is a strong candidate for DFIG replacement [10-15].

In this work, a perturbation observer-based SMC of CDFIG associated with MC for optimal power extraction is proposed. To govern the active and reactive power injection to the electrical power grid via the WPGS, controllers with dynamical variable structure and adaptive control are applied. The purpose of this of this technology is to achieve chatter-free fractional order SMC as well as decoupled active and reactive power regulation in the stator. The three-phase MC is among the recent generations of the conventional direct-power AC-AC converter. There are nine bidirectional switches in total. The MC has various advantages over traditional AC-AC converter topologies, including the ability to reverse the power flow via the MC, i.e. it can operate in all four quadrants of the voltage-current plane, allowing it to operate with both modes, either with a generator or with a motor. Assuming hypothetical zero-loss switches, the instantaneous power input should be equalized to the power output due to the lack of energy storage devices. On the other hand, it is not required that the input and output reactive powers are in agreement. On the MC inputs and depending on the load to be supplied, it is possible to modify and preset the phase angle between the voltages and currents, which does not have to be the same as the output [16-18].

The extensive research on MC began with [19]. This method was founded on a duty-cycle matrix technics and it is introduced using time domain quantities. SVM is a modulation solution, based on the representation of space vectors in complex space [20-22]. Direct SVM (DSVM) [23] and Indirect SVM (ISVM) [24] are two techniques that have been developed. The ISVM approach is investigated in this study. ISVM is a modulation approach that divides the input current and output voltage control into two phases. To regulate the entire MC, two transfer matrices are required, and the well-known SVM approach is used for the rectifying and inverting stages respectively [14].

Corresponding author: Amar Maafa

## II. DESCRIPTION OF THE WPGS

The direct MC is made up of 9 bidirectional switches that allow any input phase a, b, c to be connected to any output phase A, B, C. According to Figure 1, each output phase contains a set of 3 switches connected to the 3 input phases. The sinusoidal output voltage with variable frequency and/or magnitude must be calculated from the input voltages defined by a constant frequency and magnitude [14]. There are 2 major limitations on how an MC may operate: (a) no short circuits are permitted in the 3 input phases and (b) no circuits with 3 output phases may ever be opened [25].

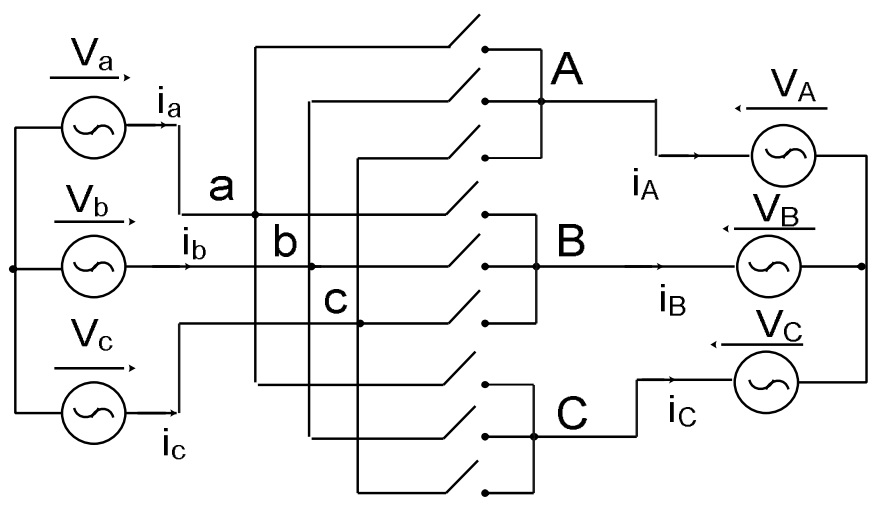

Fig. 1. Matrix converter topology.

### A. Space Vector Modulation (SVM)

The MC is represented as a two-level converter connected by an imaginary DC link in order to employ the SVPWM: the first level acts as an input rectifier with the current link and the second level acts as an output voltage source inverter [17, 26]. For further information, the interested reader is directed to [2, 21-27].

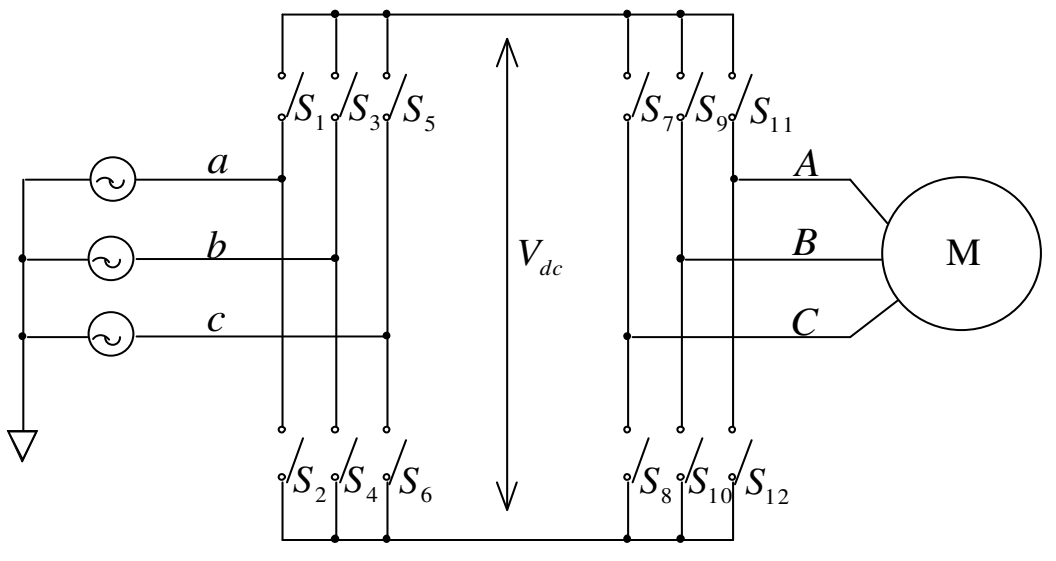

Fig. 2. Power circuit topology of the 3×3 IMC.

### B. SVM for the Rectifier Stage

The following transformation is used to describe the input current space vector $I_{in}$ as a space vector:

$$I_{in} = \frac{2}{3}(I_a + I_b e^{j\frac{2\pi}{3}} + I_c e^{-j\frac{2\pi}{3}}) \quad (1)$$

The neighboring switching vectors $I_\gamma$ and $I_\delta$ are multiplied with the corresponding duty cycles $d_\gamma$ and $d_\delta$ to create $I^*$.

$$I^* = d_\gamma I_\gamma + d_\delta I_\delta \quad (2)$$

Figure 3 shows the reference current vector synthesis.

### C. SVM For the Inverter Stage

The output voltage vector is represented as:

$$V_{out} = \frac{2}{3}(V_A + V_B e^{j\frac{2\pi}{3}} + V_C e^{-j\frac{2\pi}{3}}) \quad (3)$$

Two consecutive switching vectors $v_\alpha$ and $v_\beta$ with duty cycles $d_\alpha$ and $d_\beta$ synthesize the reference output voltage space vector as follows:

$$V^* = d_\alpha v_\alpha + d_\beta I_\beta \quad (4)$$

The active vectors' duty cycle is computed as:

$$\begin{cases} d_\alpha = \frac{T_\alpha}{T_s} = m_v \sin(\frac{\pi}{3} - \theta_v) \\ d_\beta = \frac{T_\beta}{T_s} = m_v \sin(\theta_v) \\ d_0 = \frac{T_0}{T_s} = 1 - (d_\alpha + d_\beta) \end{cases} \quad (5)$$

where $\theta_v$ represents the angle between the reference voltage vector and the actual hexagon sector. The appropriate voltage transmission ratio is determined by the voltage adjustment indicator denoted as $m_V$ [28]. In this application, for all MC bidirectional switches, it is necessary to create a single modulation mechanism from the two separate space vector modulations.

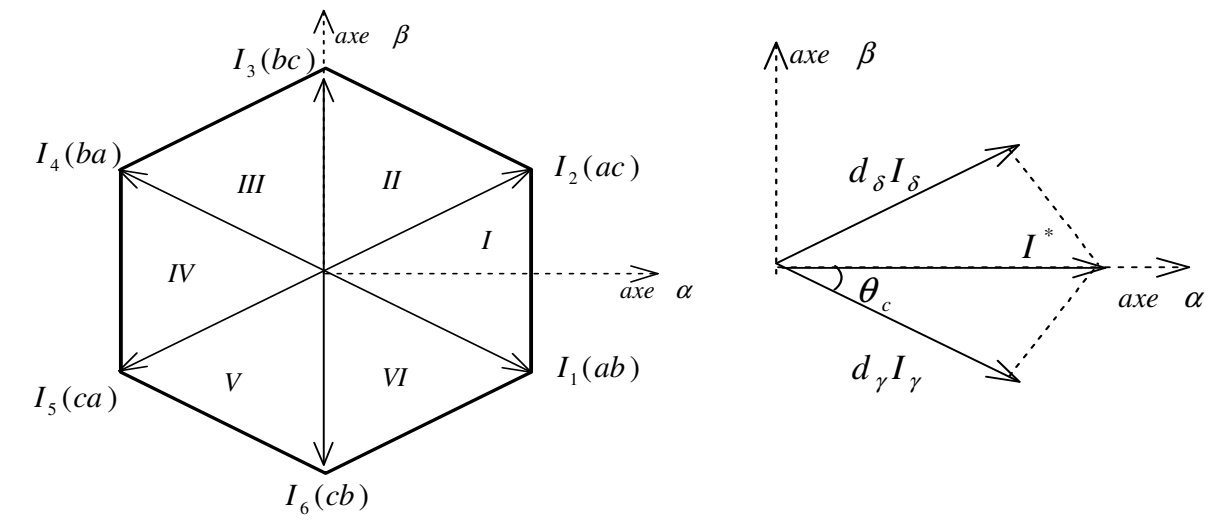

Fig. 3. Reference current vector synthesis.

### D. Model of CDFIG

The CDFIG consists of 2 induction generators, with p1 and p2 being their pole-pairs, connected in cascade to remove the brushes and copper rings in the Conventional Doubly Fed Induction Machine (DFIM) [2, 13, 29]. Figure 4 shows the mechanical and electric coupling of two wound rotors. In a synchronously rotating d–q frame, the mathematical model of the electrical machine is detailed in [2]. The purpose of the choosing of a d–q reference frame where the d-axis is in agreement with the first stator flux is to have its quadrature component equal to zero [18].

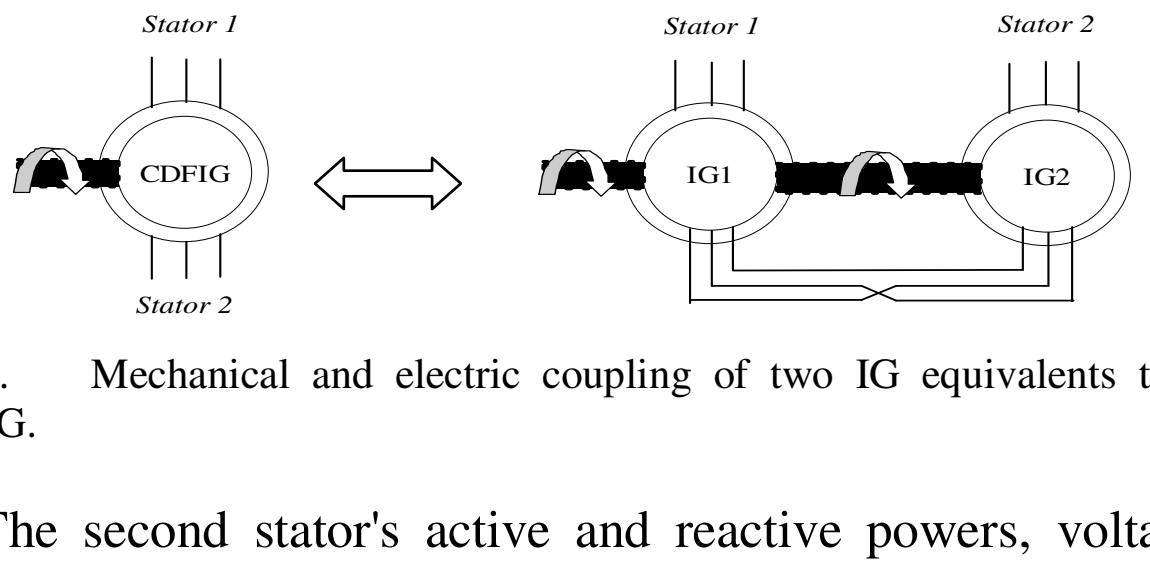

Fig. 4. Mechanical and electric coupling of two IG equivalents to the CDFIG.

The second stator's active and reactive powers, voltages, and currents can be written in accordance with the rotor currents as follows:

$$\begin{cases} v_{ds2} = R_{s2} i_{ds2} + (L_{s2} - C.L_{m2}) \frac{di_{ds2}}{dt} \\ \quad -s.\omega_s (L_{s2} - C.L_{m2}).i_{qs2} \\ v_{qs2} = R_{s2} i_{qs2} + (L_{s2} - C.L_{m2}) \frac{di_{qs2}}{dt} \\ +s.\omega_s (L_{s2} - C.L_{m2}).i_{ds2} + s.\frac{L_{m1} V_s}{L_{s1}} \end{cases} \quad (6)$$

$$\begin{cases} P_{s1} = -C.V_s \frac{L_{m1}}{L_{s1}} i_{qs2} \\ Q_{s1} = \frac{V_s^2}{\omega_s . L_{s1}} \left(1 + \frac{C.L_{m1}^2}{L_{s1}.L_{m2}}\right) - C.V_s \frac{L_{m1}}{L_{s1}} i_{ds2} \end{cases} \quad (7)$$

where $s = (\omega_s - (p_1 + p_2).\Omega_r)/\omega_s$ and $C = L_{m2}/(L_{r1} + L_{r2} - \frac{L_{m1}^2}{L_{s1}})$. The sliding surfaces $S(i_{ds2})$, $S(i_{qs2})$ relative to stator currents are defined as:

$$\begin{cases} S(i_{qs2}) = i_{qs2_ref} - i_{qs2} \\ S(i_{ds2}) = i_{ds2_ref} - i_{ds2} \end{cases} \quad (8)$$

For the chosen variables to converge to their reference values, both surfaces must be zero:

$$\begin{cases} \dot{S}(P) = \frac{d}{dt}(i_{qs2_ref} - i_{qs2}) = 0 \\ \dot{S}(Q) = \frac{d}{dt}(i_{ds2_ref} - i_{ds2}) = 0 \end{cases} \quad (9)$$

The control algorithm is defined by the relations:

$$\begin{cases} \mathrm{V_{qs2} = V_{qs2_eq} + V_{qs2_n}} \\ \mathrm{V_{ds2} = V_{ds2_eq} + V_{ds2_n}} \end{cases} \quad (10)$$

$$\begin{cases} V_{qs2-eq} = R_{s2} i_{qs2} - L_{s1} \frac{L_{s2} - CL_{m2}}{CV_s L_{m1}} \dot{P}_{s1_{ref}} \\ -s\omega_{s1}(L_{s2} - CL_{m2}) i_{ds2} + \frac{L_{m1}}{L_{s1}} CsV_s \\ V_{qs2-n} = K_1 (L_{s2} - CL_{m2})\, sgn(S(i_{qs})) \end{cases} \quad (11)$$

$$\begin{cases} V_{qs2_{eq}} = R_{s2} i_{qs2} - L_{s1} \frac{L_{s2} - CL_{m2}}{CV_s L_{m1}} \dot{P}_{s1_{ref}} \\ -s\omega_{s1}(L_{s2} - CL_{m2}) i_{ds2} + \frac{L_{m1}}{L_{s1}} CsV_s \\ V_{qs2-n} = K_1 (L_{s2} - CL_{m2})\, sgn(S(i_{qs})) \end{cases} \quad (12)$$

where $V_{qs2\text{-}eq}$, $V_{ds2\text{-}eq}$ are the equivalent control voltages used to maintain the variable to be controlled on the sliding surface, $V_{qs2\text{-}n}$ , $V_{ds2\text{-}n}$ are switching control voltages used to check the convergence condition because the parameters of the system are imprecise. An SMC block diagram applied to the CDFIG is schematized in Figure 6 according to (11)-(12). The goal is to capture the maximum aerodynamic power without exceeding the allowable power of the converter. Furthermore, the pitch angle is raised to lose aerodynamic power when the power converter's maximum rating is reached [2]. The SMC of CDFIG with MPPT using MC is shown in Figure 8.

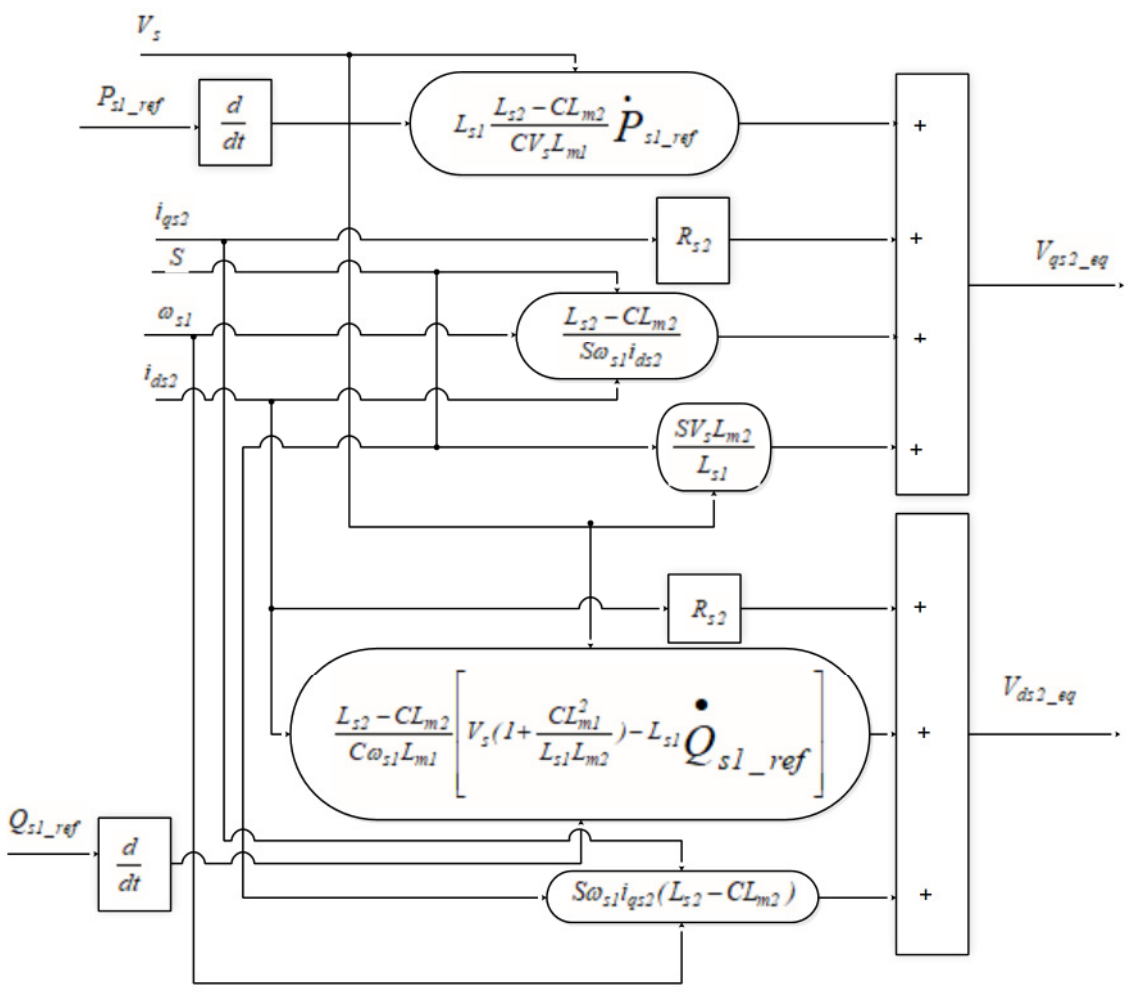

Fig. 5. The equivalent control voltage's structure.

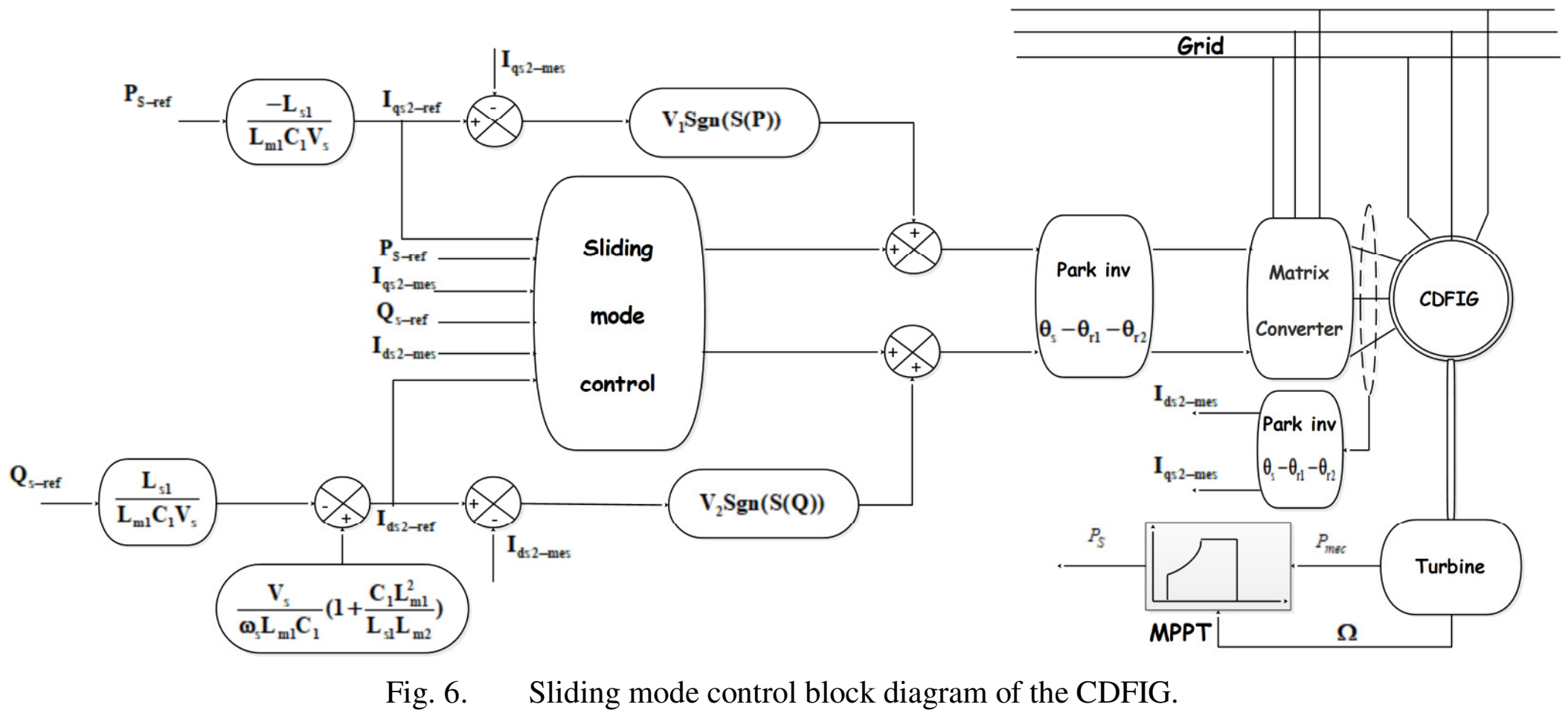

Fig. 6. Sliding mode control block diagram of the CDFIG.

## III. SIMULATION RESULTS AND DISCUSSION

The proposed WPGS was realized in Matlab/Simulink environment as presented in Figure 7, according to the control block diagram illustrated in Figure 6. We employed the DFIG parameters listed in [2]. In this section, the simulation of a direct grid connection of CDFIG through the first stator that is controlled through its second stator winding via AC/AC direct converter is presented. The considered system is controlled with the purpose to generate maximum energy while minimizing loads.

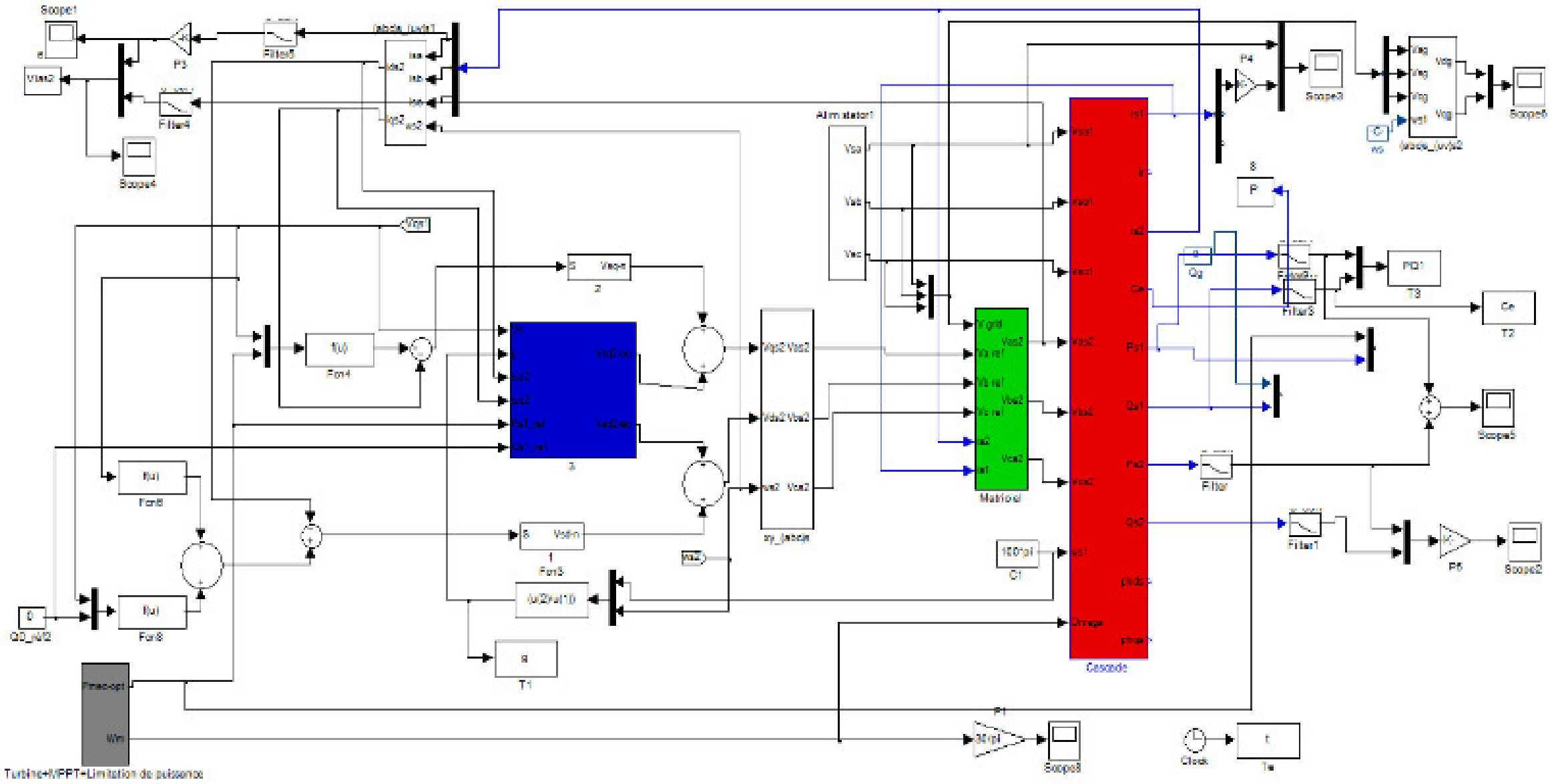

Fig. 7. The proposed WPGS model in Simulink.

Figure 8 gives the wind speed variance applied to the WPGS. Figure 9 shows the power coefficient and Figure 10 gives the corresponding pitch angle. The pitch angle of the WT is controlled to ensure that it work as hard as possible with maximum power coefficient.

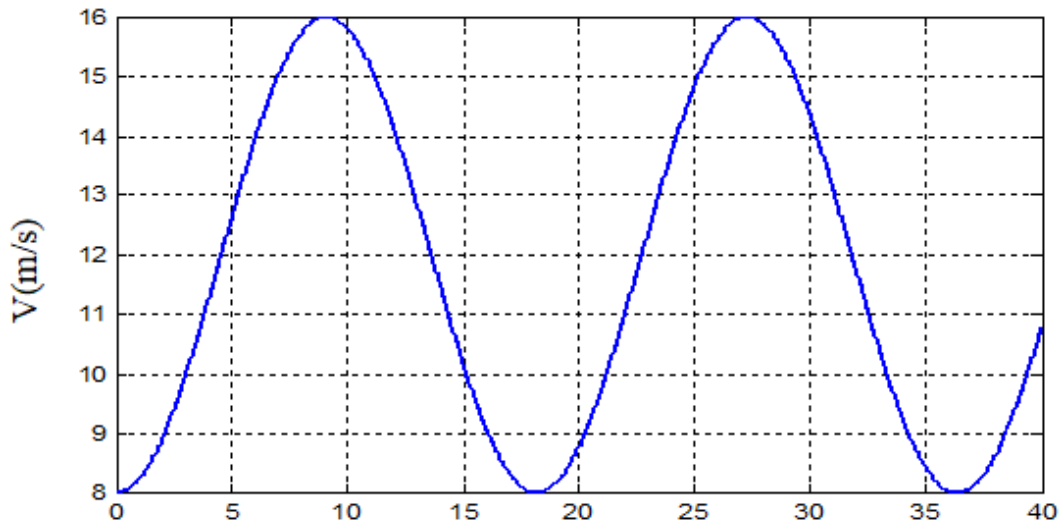

Fig. 8. Wind speed profile.

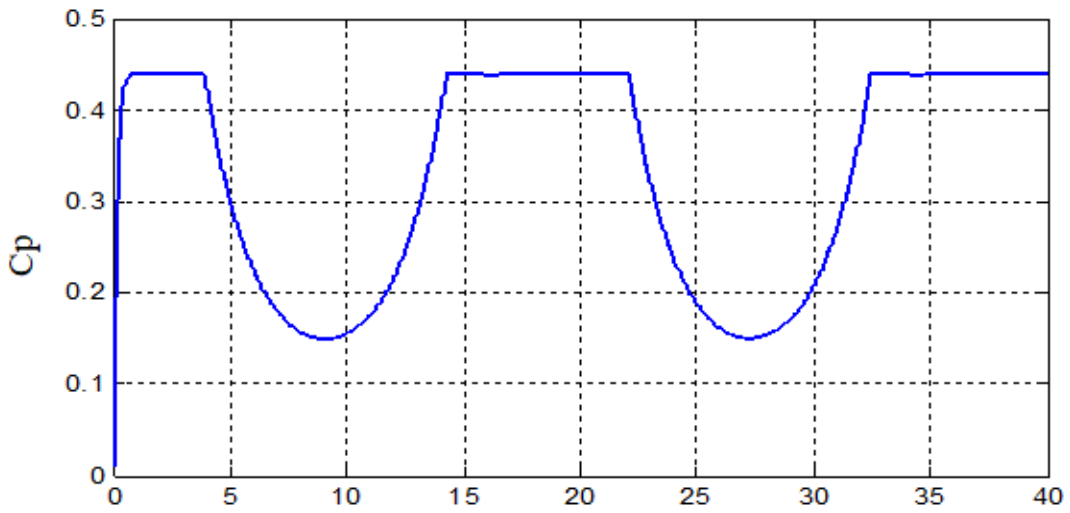

Fig. 9. Power coefficient

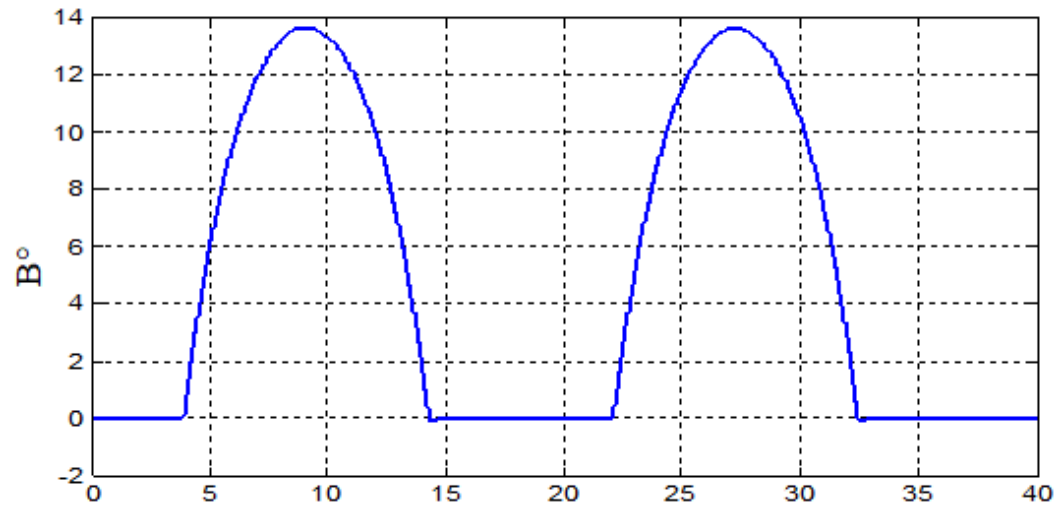

Fig. 10. Pitch angle.

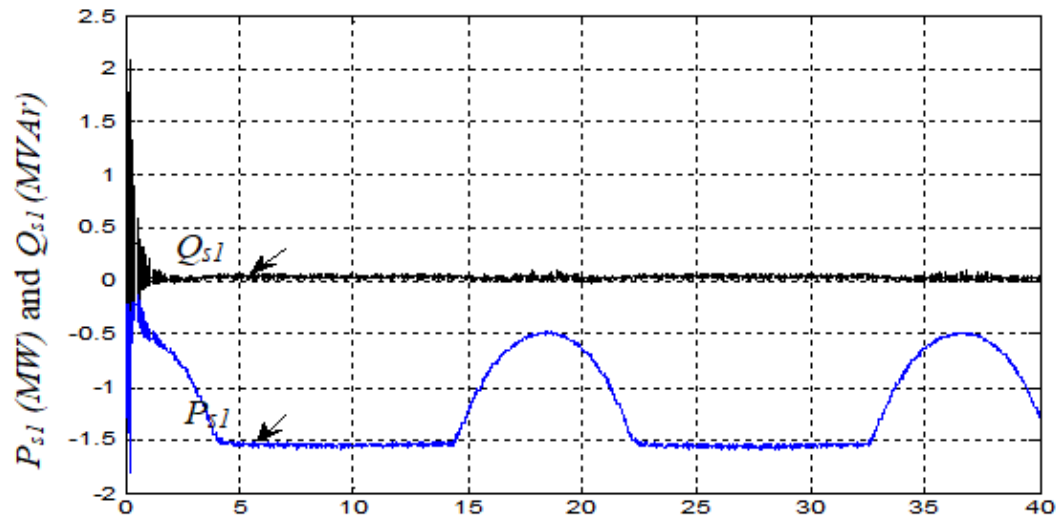

Fig. 11. First stator active and reactive powers.

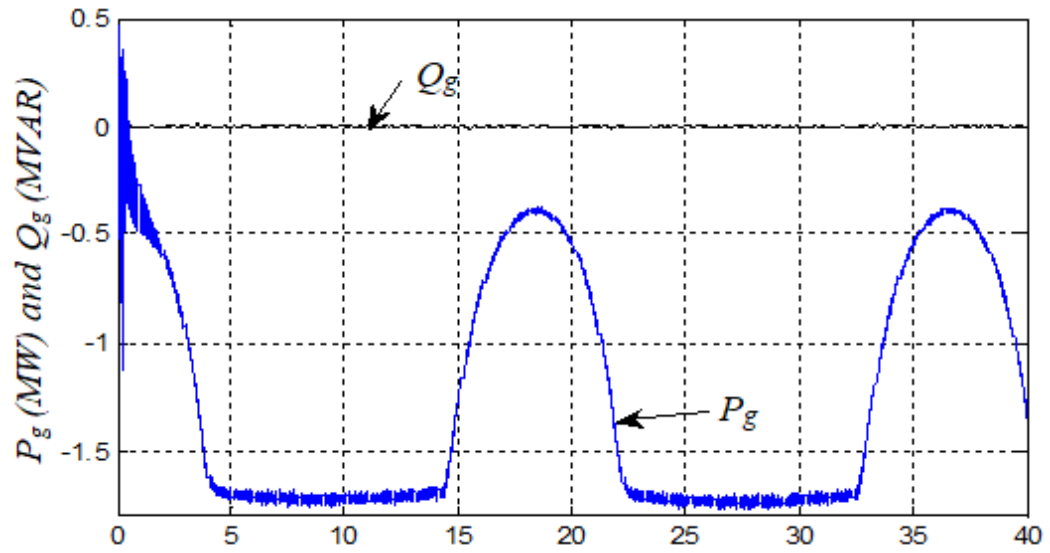

Fig. 12. Grid active and reactive powers.

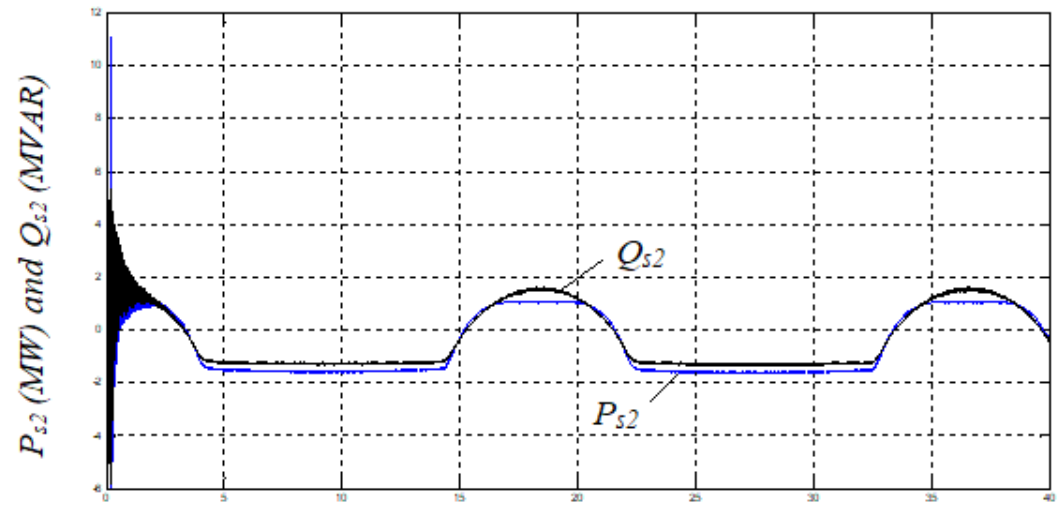

Fig. 13. Second stator active and reactive powers.

The first stator's active and reactive powers are given in Figure 11. We can see clearly that for the first stator, both active and reactive powers are in agreement with their references. The grid active and reactive powers are shown in Figure 12 and the second stator active and reactive powers are shown in Figure 13. In a hyper-synchronous operation case, it is clearly evident that a portion of the active power exceeding 1.5MW is transmitted to the grid from the second stator side of the CDFIG. The first stator's voltage and current curves are illustrated in Figure 14. As we can see, the sinusoidal current and voltage are in phase opposition as illustrated in Figure 15.

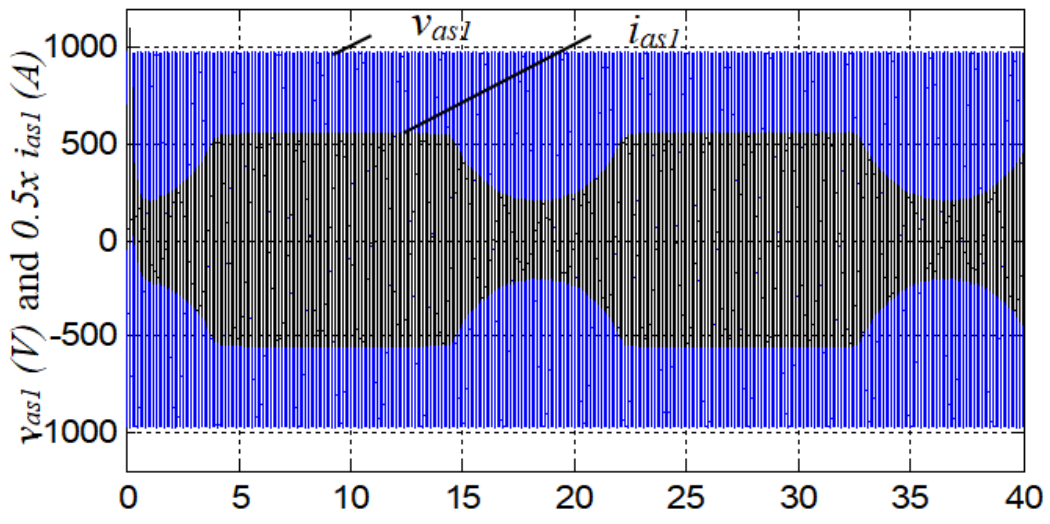

Fig. 14. First stator voltage and current.

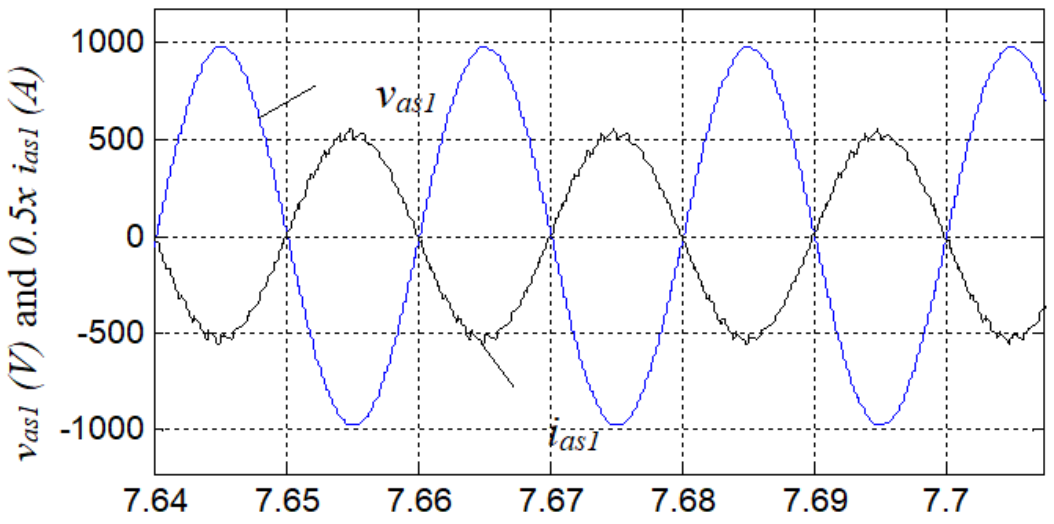

Fig. 15. Zoom first stator voltage and current.

The voltage and current curves of the second stator are depicted in Figure 16. We can conclude that according to the slip $s$ value, the frequency of the voltage and the current change, as depicted in Figures 17 and 8. Furthermore, in the case of synchronous operation mode, the slip is equal to zero and the voltage and current of the second stator have continuous forms as illustrated in Figure 19.

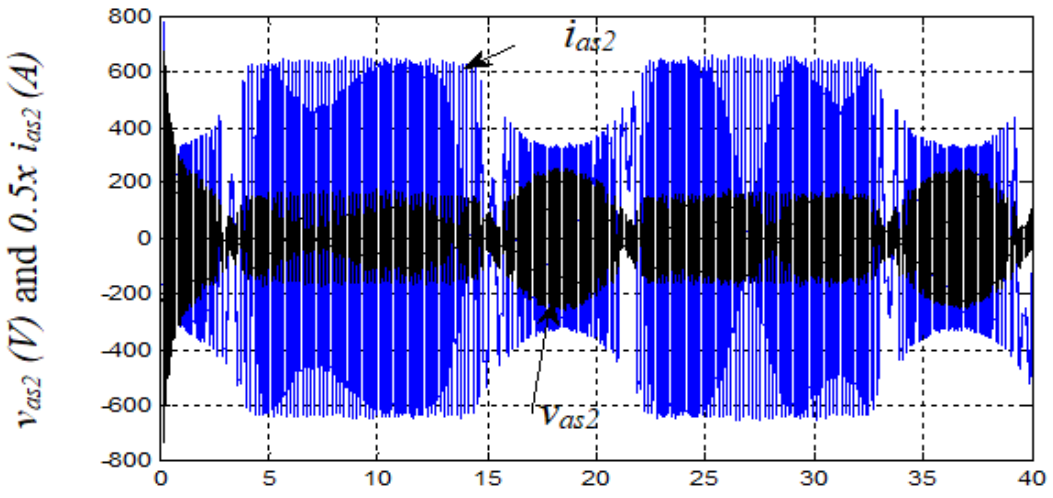

Fig. 16. Second stator voltage and current.

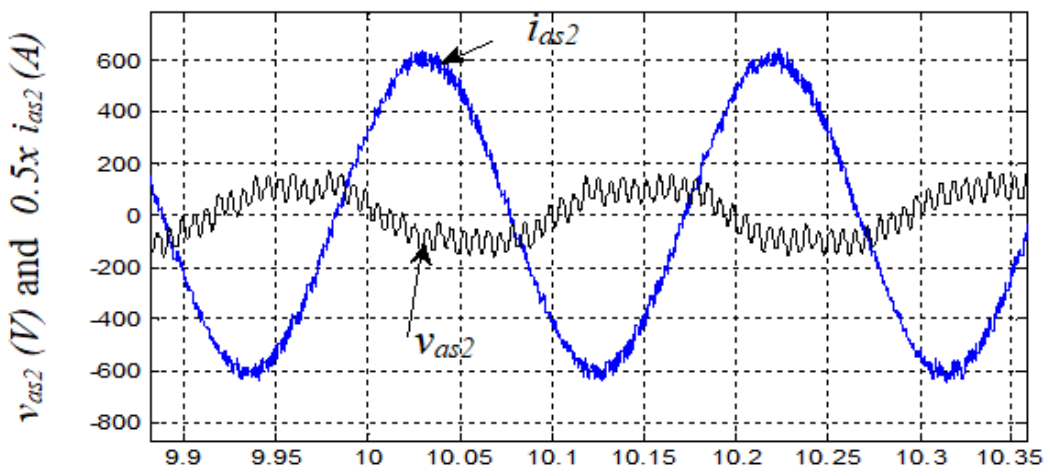

Fig. 17. Zoom second stator voltage and current ($S > 0$).

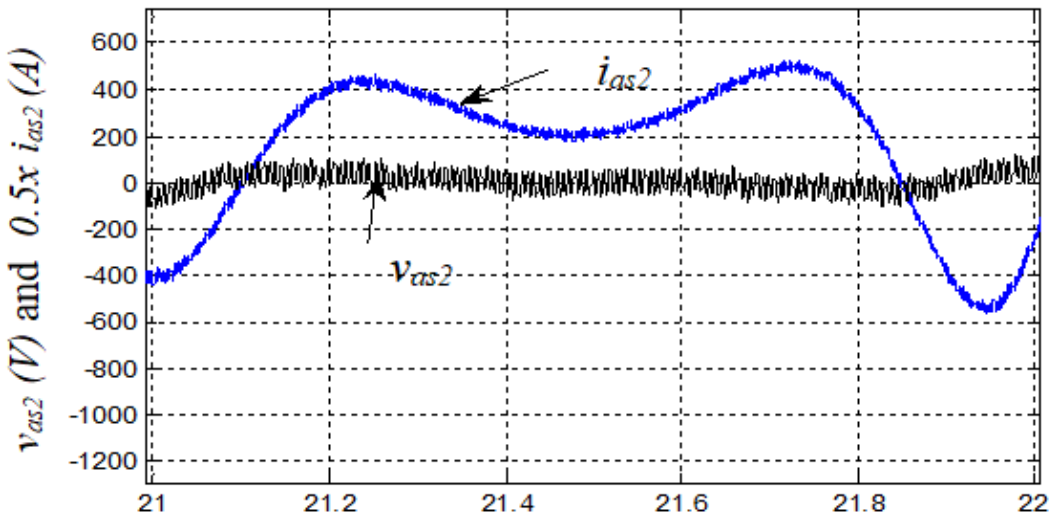

Fig. 18. Zoom second stator voltage and current ($S = 0$).

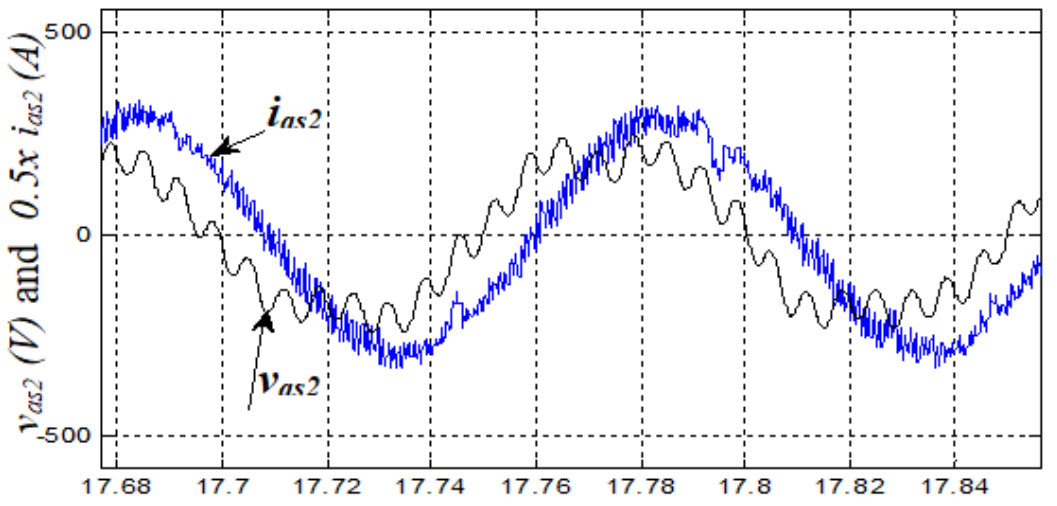

Fig. 19. Zoom second stator voltage and current ($S < 0$).

Many studies have dealt with DFIG control issues for several applications and many approaches have been proposed [30-35]. A very interested work is proposed in [30], where the authors propose SMC and an update of SMC, called terminal SMC to control the output voltage of CDFIG. Authors in [31] present a grid synchronization method founded on SMC for sub–and super–synchronous mode. A promising technique called super twisting sliding mode control was applied in [32-34] with the purpose of power control. All these works ignore pitch control and the CDFIG is not supplied via an MC. In contrast, authors in [35] use SMC with an MC but for a motor application.

## IV. Conclusion

In this paper, an SMC for indirect field-oriented CDFIG associated with MC was proposed. The suggested SMC is intended to lower chattering phenomena levels, and this has been confirmed by the simulation results of a nonlinear controlled system. The recommended MC was simulated, and the results demonstrate that it is capable of controlling the secondary side voltage and frequency even in an operational situation with bidirectional power flow. The goal of the pitch controller is to capture the maximum aerodynamic power without exceeding the power allowable of the converter. The results of the simulation showed that the active and reactive powers closely match their reference values. A near unity input power factor has been achieved by carefully managing MC's input power factor. The lack of any energy storage element in MC preserves and prolongs the life of the proposed structure. The wind speed is applied in such a way, as to have the three operating modes of the WT, namely synchronous, hyper-synchronous, and hypo-synchronous. The results obtained in this study can be found with DFIG instead of CDFIG, but CDFIG has the advantages of reducing the maintenance cost, eliminating the contacts between copper rings and brushes, and reducing the size of the gearbox in the WPGS.

## References

[1] H. Mellah, S. Arslan, H. Sahraoui, K. E. Hemsas, and S. Kamel, "The Effect of Stator Inter-Turn Short-Circuit Fault on DFIG Performance Using FEM," *Engineering, Technology & Applied Science Research*, vol. 12, no. 3, pp. 8688–8693, Jun. 2022, https://doi.org/10.48084/etasr.4923.

[2] A. Maafa, D. Aouzellag, K. Ghedamsi, and R. Abdessemed, "Cascaded doubly fed induction generator with variable pitch control system," *Revue Roumaine des Sciences Techniques*, vol. 61, no. 4, pp. 361–366, 2016.

[3] P. D. Chung, "Smoothing the Power Output of a Wind Turbine Group with a Compensation Strategy of Power Variation," *Engineering, Technology & Applied Science Research*, vol. 11, no. 4, pp. 7343–7348, Aug. 2021, https://doi.org/10.48084/etasr.4234.

[4] F. Poitiers, T. Bouaouiche, and M. Machmoum, "Advanced control of a doubly-fed induction generator for wind energy conversion," *Electric Power Systems Research*, vol. 79, no. 7, pp. 1085–1096, Jul. 2009, https://doi.org/10.1016/j.epsr.2009.01.007.

[5] D. Aouzellag, K. Ghedamsi, and E. M. Berkouk, "Network power flux control of a wind generator," *Renewable Energy*, vol. 34, no. 3, pp. 615–622, Mar. 2009, https://doi.org/10.1016/j.renene.2008.05.049.

[6] M. Adjoudj, M. Abid, A. Aissaoui, H. Bounoua, and Y. Ramdani, "Sliding mode control of a doubly fed induction generator for wind turbines," *Revue Roumaine des Sciences Techniques*, vol. 51, no. 1, pp. 15–24, 2011.

[7] K. Ghedamsi, D. Aouzellag, and E. M. Berkouk, "Control of wind generator associated to a flywheel energy storage system," *Renewable Energy*, vol. 33, no. 9, pp. 2145–2156, Sep. 2008, https://doi.org/10.1016/j.renene.2007.12.009.

[8] A. Djoudi, S. Bacha, H. Iman-Eini, and T. Rekioua, "Sliding mode control of DFIG powers in the case of unknown flux and rotor currents with reduced switching frequency," *International Journal of Electrical Power & Energy Systems*, vol. 96, pp. 347–356, Mar. 2018, https://doi.org/10.1016/j.ijepes.2017.10.009.

[9] M. Rahimi, "Improvement of energy conversion efficiency and damping of wind turbine response in grid connected DFIG based wind turbines," *International Journal of Electrical Power & Energy Systems*, vol. 95, pp. 11–25, Feb. 2018, https://doi.org/10.1016/j.ijepes.2017.08.005.

[10] R. Cardenas, R. Pena, P. Wheeler, J. Clare, A. Munoz, and A. Sureda, "Control of a wind generation system based on a Brushless Doubly-Fed Induction Generator fed by a matrix converter," *Electric Power Systems Research*, vol. 103, pp. 49–60, Oct. 2013, https://doi.org/10.1016/j.epsr.2013.04.006.

[11] M. El Achkar, R. Mbayed, G. Salloum, S. Le Ballois, and E. Monmasson, "Generic study of the power capability of a cascaded doubly fed induction machine," *International Journal of Electrical Power & Energy Systems*, vol. 86, pp. 61–70, Mar. 2017, https://doi.org/10.1016/j.ijepes.2016.09.011.

[12] L. Bouchaoui, K. Hemsas, H. Mellah, and S. Benlahneche, "Power transformer faults diagnosis using undestructive methods (Roger and IEC) and artificial neural network for dissolved gas analysis applied on the functional transformer in the Algerian north-eastern: a comparative study," *Electrical Engineering & Electromechanics*, no. 4, pp. 3–11, Apr. 2021, https://doi.org/10.20998/2074-272X.2021.4.01.

[13] K. Protsenko and D. Xu, "Modeling and Control of Brushless Doubly-Fed Induction Generators in Wind Energy Applications," *IEEE Transactions on Power Electronics*, vol. 23, no. 3, pp. 1191–1197, Feb. 2008, https://doi.org/10.1109/TPEL.2008.921187.

[14] M. Moazen, R. Kazemzadeh, and M.-R. Azizian, "Mathematical modeling and analysis of brushless doubly fed reluctance generator under unbalanced grid voltage condition," *International Journal of Electrical Power & Energy Systems*, vol. 83, pp. 547–559, Dec. 2016, https://doi.org/10.1016/j.ijepes.2016.04.050.

[15] M. Adamowicz and R. Strzelecki, "Cascaded doubly fed induction generator for mini and micro power plants connected to grid," in *13th International Power Electronics and Motion Control Conference*, Poznan, Poland, Sep. 2008, pp. 1729–1733, https://doi.org/10.1109/EPEPEMC.2008.4635516.

[16] N. Taib, B. Metidji, and T. Rekioua, "Performance and efficiency control enhancement of wind power generation system based on DFIG using three-level sparse matrix converter," *International Journal of Electrical Power & Energy Systems*, vol. 53, pp. 287–296, Dec. 2013, https://doi.org/10.1016/j.ijepes.2013.05.019.

[17] N. Mohan, T. M. Undeland, and W. P. Robbins, *Power Electronics: Converters, Applications, and Design*, 3rd ed. New York, NY, USA: Wiley, 2002.

[18] A. Djahbar, B. Benziane, and A. Zegaoui, "A Novel Modulation Method for Multilevel Matrix Converter," *Energy Procedia*, vol. 50, pp. 988–998, Jan. 2014, https://doi.org/10.1016/j.egypro.2014.06.118.

[19] A. Alesina and M. G. B. Venturini, "Analysis and design of optimum-amplitude nine-switch direct AC-AC converters," *IEEE Transactions on Power Electronics*, vol. 4, no. 1, pp. 101–112, Jan. 1989, https://doi.org/10.1109/63.21879.

[20] V. T. Ha, P. T. Giang, and V. H. Phuong, "T-Type Multi-Inverter Application for Traction Motor Control," *Engineering, Technology & Applied Science Research*, vol. 12, no. 2, pp. 8321–8327, Apr. 2022, https://doi.org/10.48084/etasr.4776.

[21] A. Benachour, E. M. Berkouk, and M. Mahmoudi, "A new direct torque control of induction machine fed by indirect matrix converter," *Revue Roumaine des Sciences Techniques*, vol. 62, pp. 25–30, Jan. 2017.

[22] F. Yue, P. W. Wheeler, and J. C. Clare, "Relationship of Modulation Schemes for Matrix Converters," in *3rd IET International Conference on Power Electronics, Machines and Drives - PEMD 2006*, Dublin, Ireland, Apr. 2006, pp. 266–270, https://doi.org/10.1049/cp:20060113.

[23] D. Casadei, G. Grandi, G. Serra, and A. Tani, "Space vector control of matrix converters with unity input power factor and sinusoidal input/output waveforms," in *Fifth European Conference on Power Electronics and Applications*, Brighton, UK, Sep. 1993, vol. 7, pp. 170–175.

[24] L. Huber and D. Borojevic, "Space vector modulated three-phase to three-phase matrix converter with input power factor correction," *IEEE Transactions on Industry Applications*, vol. 31, no. 6, pp. 1234–1246, Aug. 1995, https://doi.org/10.1109/28.475693.

[25] P. Bunpakdee, J. Theeranan, and C. Jeraputra, "Experimental Validation of Vector Control of a Matrix Converter Fed Induction Generator for Renewable Energy Sources," *Procedia Computer Science*, vol. 86, pp. 397–400, Jan. 2016, https://doi.org/10.1016/j.procs.2016.05.043.

[26] S. M. Dabour, A. S. Abdel-Khalik, S. Ahmed, A. M. Massoud, and S. M. Allam, "Common-mode voltage reduction for space vector modulated three- to five-phase indirect matrix converter," *International Journal of Electrical Power & Energy Systems*, vol. 95, pp. 266–274, Feb. 2018, https://doi.org/10.1016/j.ijepes.2017.08.020.

[27] D. Varajao and R. E. Araujo, "Modulation Methods for Direct and Indirect Matrix Converters: A Review," *Electronics*, vol. 10, no. 7, Jan. 2021, Art. no. 812, https://doi.org/10.3390/electronics10070812.

[28] A. Abdelrehim, S. D. Erfan, M. El-Habrouk, and K. H. Youssef, "Indirect 3D-Space Vector Modulation for Matrix Converter," *Engineering, Technology & Applied Science Research*, vol. 8, no. 2, pp. 2847–2852, Apr. 2018, https://doi.org/10.48084/etasr.2007.

[29] A. Maafa, D. Aouzellag, K. Ghedamsi, and R. Abdessemed, *Modélisation et contrôle en puissance d'une cascade de deux machines asynchrones doublement alimentées (CDFIG).* Batna, Algeria: University of Batna 2, 2010.

[30] H. Zahedi Abdolhadi, G. Arab Markadeh, and S. Taghipour Boroujeni, "Sliding Mode and Terminal Sliding Mode Control of Cascaded Doubly Fed Induction Generator," *Iranian Journal of Electrical and Electronic Engineering*, vol. 17, no. 3, pp. 1955–1955, Sep. 2021, https://doi.org/10.22068/IJEEE.17.3.1955.

[31] X. Yan and M. Cheng, "A Robust Grid Synchronization Method for Cascaded Brushless Doubly Fed Induction Generator," in *22nd International Conference on Electrical Machines and Systems*, Harbin, China, Aug. 2019, pp. 1–6, https://doi.org/10.1109/ICEMS.2019.8922507.

[32] X. Yan and M. Cheng, "A Robustness-Improved Control Method Based on ST-SMC for Cascaded Brushless Doubly Fed Induction Generator,"

*IEEE Transactions on Industrial Electronics*, vol. 68, no. 8, pp. 7061–7071, Dec. 2021, https://doi.org/10.1109/TIE.2020.3007087.

[33] R. Sadeghi, S. M. Madani, M. Ataei, M. R. Agha Kashkooli, and S. Ademi, "Super-Twisting Sliding Mode Direct Power Control of a Brushless Doubly Fed Induction Generator," *IEEE Transactions on Industrial Electronics*, vol. 65, no. 11, pp. 9147–9156, Aug. 2018, https://doi.org/10.1109/TIE.2018.2818672.

[34] O. Moussa, R. Abdessemed, and S. Benaggoune, "Super-twisting sliding mode control for brushless doubly fed induction generator based on WECS," *International Journal of System Assurance Engineering and Management*, vol. 10, no. 5, pp. 1145–1157, Oct. 2019, https://doi.org/10.1007/s13198-019-00844-3.

[35] H. Wang *et al.*, "A Cascade PI-SMC Method for Matrix Converter-Fed BDFIM Drives," *IEEE Transactions on Transportation Electrification*, vol. 7, no. 4, pp. 2541–2550, Sep. 2021, https://doi.org/10.1109/TTE.2021.3061742.